# Imaging Quantum Interference in Stadium-Shaped Monolayer and Bilayer Graphene Quantum Dots

Zhehao Ge[6, †], Dillon Wong[1,†], Juwon Lee[1], Frederic Joucken[6], Eberth A. Quezada-Lopez[6], Salman Kahn[1], Hsin-Zon Tsai[1], Takashi Taniguchi[4], Kenji Watanabe[5], Feng Wang[1,2,3], Alex Zettl[1,2,3], Michael F. Crommie[1,2,3, *], Jairo Velasco Jr.[6,*]

[1]*Department of Physics, University of California, Berkeley, California 94720, USA*
[2]*Materials Sciences Division, Lawrence Berkeley National Laboratory, Berkeley, California 94720, USA*
[3]*Kavli Energy NanoSciences Institute at the University of California, Berkeley and the Lawrence Berkeley National Laboratory, Berkeley, California 94720, USA*
[4]*International Center for Materials Nanoarchitectonics, National Institute for Materials Science, 1-1 Namiki, Tsukuba, 305-0044, Japan*
[5]*Research Center for Functional Materials, National Institute for Materials Science, 1-1 Namiki, Tsukuba, 305-0044, Japan*
[6] *Department of Physics, University of California, Santa Cruz, California 95064, USA*
[†] *These authors contribute equally to this manuscript.*
* *Email: jvelasc5@ucsc.edu & crommie@berkeley.edu*

**Abstract**

Experimental realization of graphene-based stadium-shaped quantum dots (QDs) have been few and incompatible with scanned probe microscopy. Yet, direct visualization of electronic states within these QDs is crucial for determining the existence of quantum chaos in these systems. We report the fabrication and characterization of electrostatically defined stadium-shaped QDs in heterostructure devices composed of monolayer graphene (MLG) and bilayer graphene (BLG). To realize a stadium-shaped QD, we utilized the tip of a scanning tunneling microscope to charge defects in a supporting hexagonal boron nitride flake. The stadium states visualized are consistent with tight-binding-based simulations, but lack clear quantum chaos signatures. The absence of quantum chaos features in MLG-based stadium QDs is attributed to the leaky nature of the confinement potential due to Klein tunneling. In contrast, for BLG-based stadium QDs (which have stronger confinement) quantum chaos is precluded by the smooth confinement potential which reduces interference and mixing between states.

The advent of pristine, exposed circular p-n junctions in monolayer graphene (MLG) [1] and Bernal-stacked bilayer graphene (BLG) [2] has enabled realization of electrostatically defined quantum dots (QDs) that are accessible to atomically-resolved scanning probe microscopy. The charge carriers of these QDs possess chirality, and so their electronic states are exotic and unlike the states in conventional semiconductor QDs [3-5]. For example, recent scanning tunneling microscopy (STM) studies of electrostatically defined MLG and BLG QDs have revealed relativistic quasibound states [6, 7], correlated states manifesting a wedding-cake-like charge density [8], and states with broken rotational symmetry and nontrivial band topology[9]. So far, the QD systems studied using this technique have all been based on circular-symmetric boundaries [6, 8-10].

Other QD symmetries, however, are possible and create new opportunities to observe novel behavior. For example, electron transport studies of conventional semiconductor-based QDs have demonstrated that circular- and square-shaped QDs host regular dynamics, while stadium-shaped QDs are described by chaotic quantum billiards [11-14]. This is the quantum mechanical analogue to the Bunimovich billiard (also known as the stadium billiard), a well-known classical chaotic system [15]. Chaotic behavior here is attributed to the nonintegrability of the classical dynamics in a stadium billiard, i.e. the existence of open orbits. Novel electronic states are expected to result from the quantum mechanical behavior of such classically chaotic systems. An example is scarred wavefunctions which are characterized by standing waves that follow semiclassical periodic orbits [13]. Extensive theoretical analysis has been performed on scarred wavefunctions [13, 16] and numerous experimental studies have imaged analogous phenomena in surface water waves [17], ultrasonic acoustic fields [18], microwave cavities [19, 20], and soap films [21]. A number of predictions for scarred wavefunctions in graphene-based systems have also been made [22, 23], but no direct visualization of electronic states in electrostatically defined nonintegrable MLG or BLG QDs has yet been reported.

Here we report the fabrication and imaging of electrostatically defined MLG and BLG QDs that are nonintegrable. The QDs were fabricated using an STM-based technique whereby electrostatic charge is injected directly into the hexagonal boron nitride (hBN) insulating layer underlying our graphene samples. QDs with stadium-shaped boundaries were synthesized by performing multiple tip injections to define each QD. Our STM measurements of the electronic wavefunctions of stadium-shaped QDs showed significant differences between MLG and BLG stadia, including interior nodal patterns and the presence of diagonal streaking, but no clear evidence of scarring phenomena was observed. Simulations of our QDs using a tight-binding based formalism reveals that the lack of scarring for the MLG stadium arises due to Klein tunneling at the stadium boundary, whereas in BLG the main culprit is the smoothness of the QD boundary. Diagonal streaking in the BLG stadium is shown to be a signature of the non-integrable nature of the QD boundary.

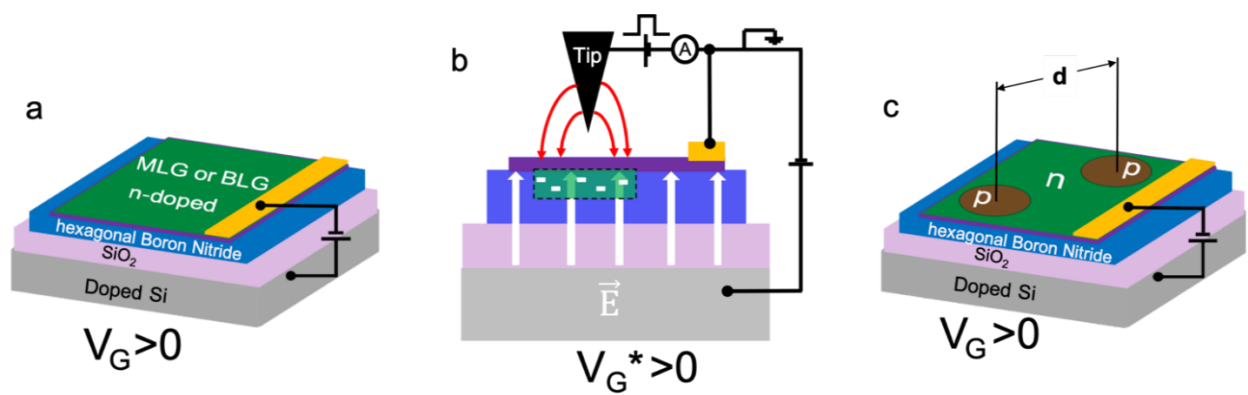

**Figure 1: Schematic for creating non-circular quantum dots (QDs) on monolayer graphene (MLG) or bilayer graphene (BLG) on hexagonal boron nitride (hBN).** **(a)** MLG or BLG/hBN heterostructure supported by SiO$_2$/Si substrate. The graphene layer is contacted by gold electrodes and a backgate voltage $V_G$ is applied to the doped Si substrate. **(b)** To fabricate multiple p-n junctions, voltage pulses are applied to the scanning tunneling microscope (STM) tip at two different locations on the MLG or BLG sample while holding $V_G^* > 0$. **(c)** Spatial control of the resulting p-n junction (i.e., QD) geometry is achieved by applying voltage pulses at different locations separated by a distance $= d$.

Figure 1 sketches our method for creating noncircular QDs in a MLG (or BLG)/hBN heterostructure supported by an SiO$_2$/Si substrate. A backgate voltage ($V_G$) applied to the doped Si substrate permits global tuning of the MLG (or BLG) doping level (Fig. 1a). Local modification of the potential is achieved by applying a 5V pulse to the STM tip for 60 secs that causes electric-field-induced excitation of defect states in hBN below the STM tip [2, 24]. This liberates defect charge that drifts towards the MLG (or BLG) under the applied $V_G$*, thus leaving behind charged ions that locally gate the MLG (or BLG) above. It is possible to create multiple QDs by simply repeating this procedure at points separated by a distance d (Fig. 1c). Additional tuning of the QD behavior is possible by subsequently modulating $V_G$, as shown in Fig. 2a. Here the potential landscape for two QDs is represented by two bell-shaped potentials having centers spaced apart by d [6, 10, 25]. Increasing $V_G$ causes a vertical shift of the potential landscape with respect to the MLG (or BLG) chemical potential ($\mu_g$), which is shown as a dashed red line in Fig. 2a. Such $V_G$ modulation changes the width of each bell-shaped potential at $\mu_g$, thus enabling modification of the QD size.

To experimentally characterize how the properties of adjacent QDs are modified by changing $V_G$, we performed spatially-resolved spectroscopic mapping of QDs fabricated using the local electric-field treatment described above. Figures 2b-d show spectroscopic characterization of two adjacent QDs fabricated in a MLG/hBN heterostructure. $d^2I/dV_S^2$ is plotted as a function of sample-tip bias along a line extending from the center of one circular QD to the outer edge of the adjacent circular QD. Each panel represents a measurement performed at a different value of $V_G$. An outer envelope is observed (blue) along with internal nodal structure (red and blue). As $V_G$ is reduced these features all shift upwards with respect to $V_S$=0, which corresponds to $\mu_g$. Because these data originate from a MLG/hBN heterostructure, the observed nodal pattern corresponds to relativistic quasibound states similar to those previously reported for electrostatically defined MLG QDs [6]. The notable bending of these states near the QD boundary reflects the influence of the biased STM tip as discussed in Quezada, et al. [25].

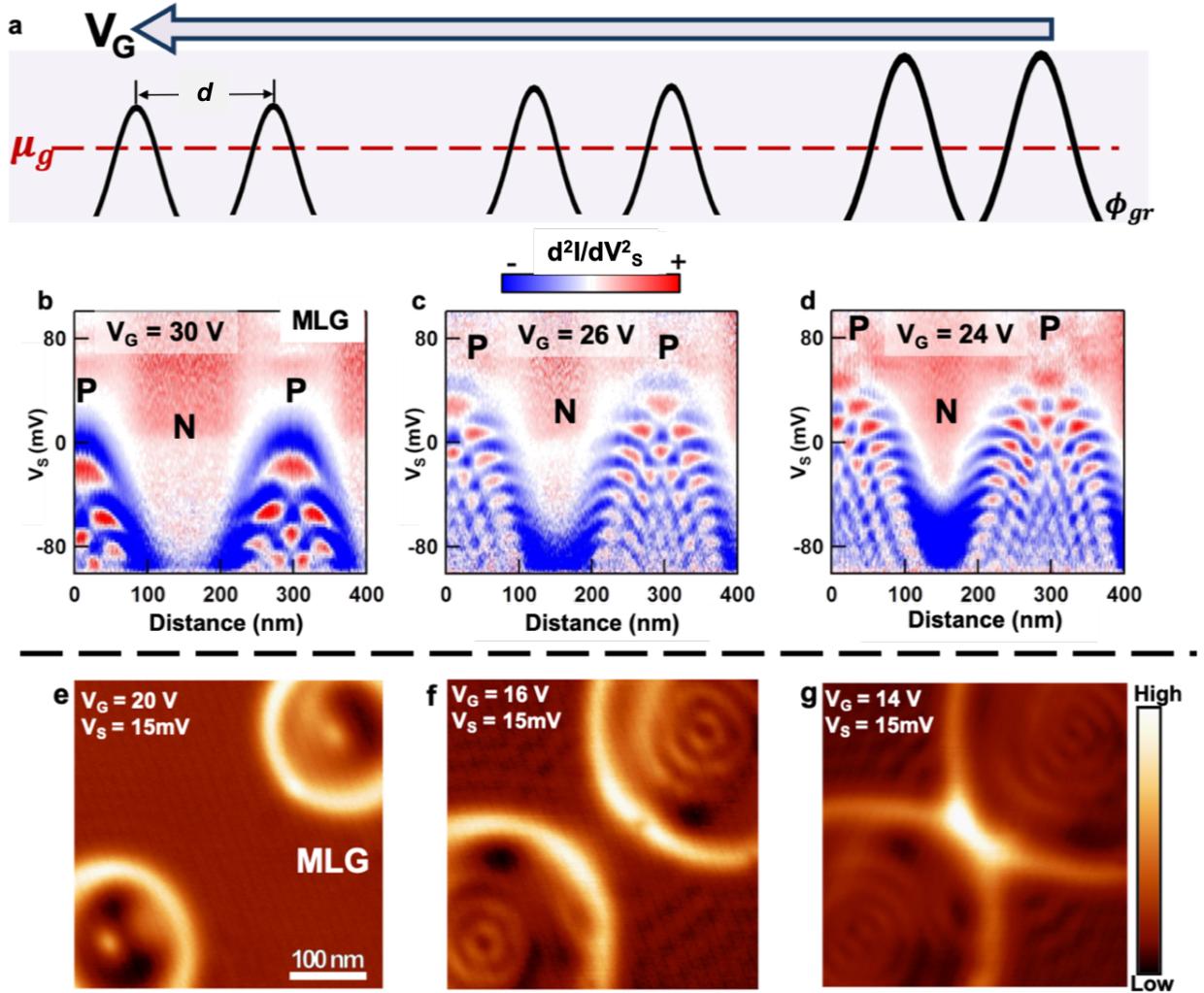

**Figure 2: Control and visualization of gate-tunable quantum dots in MLG. (a)** Sketch shows how modulation of $V_G$ allows the QD electrostatic potential to be shifted with respect to the MLG chemical potential, $\mu_g$. **(b)** $d^2I/dV_S^2$ measured as a function of sample bias ($V_S$) and distance from the center of one QD to the outer edge of a second adjacent QD ($V_G = 30$ V, $V_S = -0.1$ V, $I = 1.0$ nA, 1 mV a.c. modulation). **(c,d)** same as (b), but with (c) $V_G = 26$ V and (d) $V_G = 24$ V. **(e)** $dI/dV_S$ map for two MLG QDs ($V_S = 15$ mV, $I = 0.5$ nA, $V_G = 20$ V, 1 mV a.c. modulation). **(f,g)** $dI/dV_S$ maps at the same location as (e), but with **(f)** $V_G = 16$V and **(g)** $V_G = 14$V. The bright bands (regions of high $dI/dV_S$) indicate the location of the QD barrier wall. As $V_G$ is decreased the circular p-doped graphene regions increase and merge.

A more complete picture of the QD shape modification and electronic states is attained by acquiring $dI/dV_S$ maps at constant sample-tip bias and different $V_G$ (Figs. 2e-g). Each $dI/dV_S$ map here displays two circular QDs with bright boundaries surrounded by outer regions having little $dI/dV_S$ intensity change (the QDs shown in Figs. 2e-g are different from those shown in Figs. 2b-d, but were acquired using a similar procedure). As the gate voltage is decreased from $V_G = 20V$ (Fig. 2e) to $V_G = 16V$ (Fig. 2f) the internal structure of the QDs is seen to include more concentric, circularly symmetric states. For example, in Fig. 2e only one central antinode is visible, while after reducing $V_G$ numerous ring-like states appear. The increase in the number of ring-like states indicates that higher-energy QD states are being probed. Decrease of the gate voltage is also seen to cause the diameter of each QD to increase, eventually causing the QDs to merge, as shown in Fig. 2g. $V_G$ modulation thus enables realization of new, noncircular QDs through the merging of circular QDs.

We have exploited this technique to fabricate noncircular MLG and BLG QDs with boundary geometries similar to nonintegrable QDs previously studied in semiconductor heterojunctions [11, 14]. This was accomplished by performing multiple tip pulses with a separation of d~100 or d~50 nm and then modulating $V_G$ to further control the QD electronic states. Figures 3a,b show constant-bias $dI/dV_S$ maps at different $V_G$ values for a noncircular MLG QD created using this procedure. The data reveal a stadium-shaped structure with internal nodal patterns that change as the gate voltage is decreased from $V_G = 16V$ (Fig. 3a) to $V_G = 10V$ (Fig. 3b). Fig. 3a exhibits a vertical dark stripe in the center of the stadium QD which evolves into a pattern of antinodes as $V_G$ is reduced (Fig. 3b). Additional characteristic features visible in Fig. 3b are two structures having trigonal symmetry that lie in each half of the stadium (outlined by two triangles).

We similarly created stadium-shaped BLG QDs as shown in Figs. 3c,d. These QDs also exhibit a vertical dark stripe at the stadium center at high gate voltage as well as trigonal patterns at the stadium ends that become visible at lower gate voltage. The trigonal patterns in the BLG QD are different than those seen in the MLG QD because the BLG patterns "point" in the same direction (see triangle outlines in Fig. 3d). The BLG stadium QD also shows streaks in $dI/dV_S$ intensity that emanate diagonally from the QD trigonal patterns and intersect with the top and bottom walls. These streaks are not seen in the MLG QD.

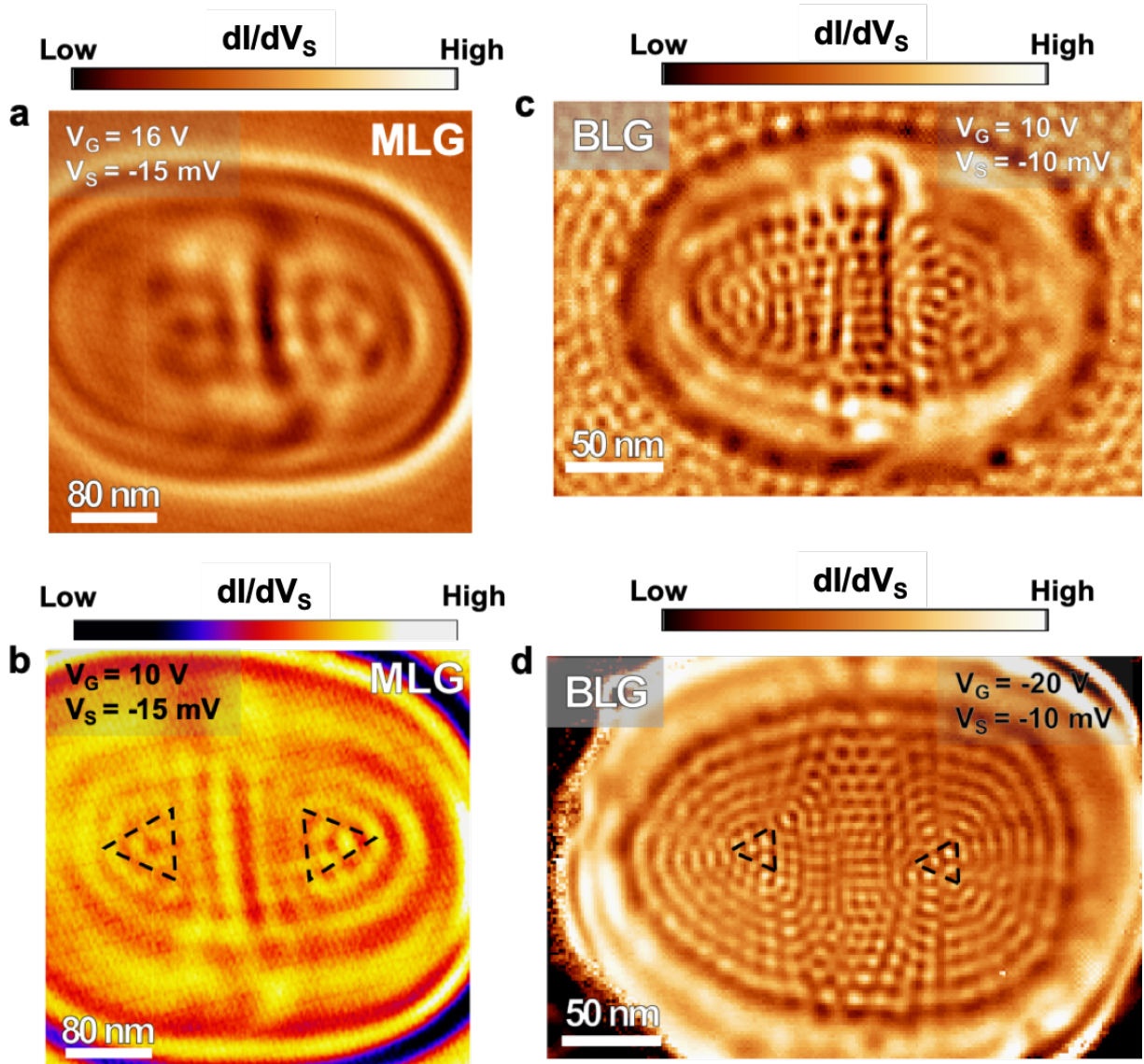

**Figure 3: Visualization of gate-tunable stadium-shaped QDs in MLG and BLG sheets.** (a) $dI/dV_S$ map for a stadium-shaped MLG QD ($V_S$ = -15 mV, $I$ = 0.5 nA, $V_G$ = 16 V). (b) $dI/dV_S$ map at the same location as (a), but with $V_G$ = 10V. Triangles outline trigonally symmetric nodal patterns. (c) $dI/dV_S$ map for a stadium-shaped BLG QD ($V_S$ = -10 mV, $I$ = 0.1 nA, $V_G$ = 10 V). (d) $dI/dV_S$ map at the same location as (c), but with $V_G$ = -20V. Triangles outline trigonally symmetric nodal patterns. As $V_G$ is decreased, the size of the stadium-shaped QDs increase and higher-energy QD states are visualized.

We now discuss the origin of the patterns in the MLG and BLG stadiums, as well as the apparent absence of scarred wavefunctions (more $dI/dV_S$ maps for MLG and BLG stadia showing the absence of scarred wavefunctions can be found in SI section 5). We first note that the dark vertical stripe seen in the central regions of both the MLG and BLG stadia can be attributed to a charge-trapping potential. This arises since the overall QD potential is created by multiple tip-pulses spaced a distance apart, and so the top of the potential takes the form of two bell-shaped structures with a dip (i.e., vertical stripe) in the middle. Decreasing $V_G$ causes the bell-shaped potentials to merge and the QD electronic state at $E_F$ to shift from the top of the potential to a lower regime where the potential landscape is more homogeneous and where the center of the stadium exhibits a more complex nodal structure.

To further understand our experimental findings, we performed numerical tight-binding-based simulations of MLG and BLG stadium-shaped QDs. A flat-bottomed potential with smooth boundaries was used for the simulations and we explored both ungapped MLG electronic structure and gapped BLG electronic structure with trigonally warped bands. These configurations are schematically depicted in Figs. 4a,b, and the resulting simulations of the stadium constant-energy $dI/dV_S$ maps (i.e., the resulting local density of states (LDOS) patterns) are shown in Figs. 4d,e. More details on the simulations can be found in Supporting Information Section 1. Similarities between the simulations and the experimental stadia are evident, and the role of the QD walls is also revealed by the calculations. For example, in the gapless MLG calculation of Fig. 4d a three-fold-symmetric pattern that is seen on each side of the stadium (triangles) that points outward and agrees with the experimental data of Fig. 3b. The simulated BLG stadium in Fig. 4e displays threefold symmetric patterns on each side that point in the same direction (see triangles) and are similar to the data seen in Fig. 3d (see SI section 2 for close-up plots). Fig. 4e also exhibits diagonal streaks that are in qualitative agreement with the experimental $dI/dV_S$ intensity of Fig. 3d.

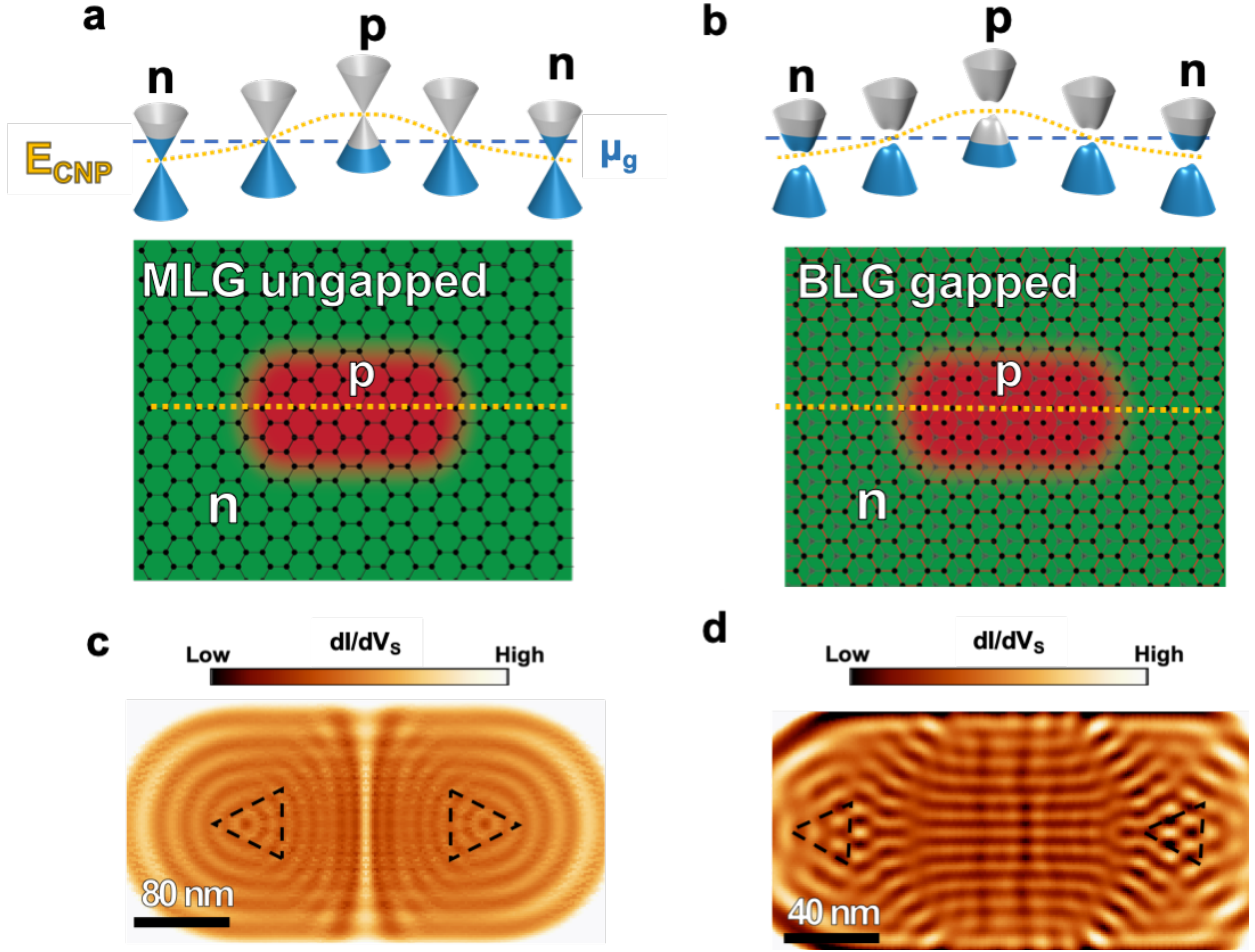

**Figure 4: Stadium QD potential profile schematic and associated simulations. (a)** Upper panel: MLG stadium QD potential profile schematic depicting the MLG bands and charge neutrality point ($E_{CNP}$) both inside and outside of the QD. Lower Panel: Schematic representation of stadium MLG QD. **(b)** Upper panel: BLG stadium QD potential profile schematic depicting the gapped and trigonally warped BLG bands and $E_{CNP}$ inside and outside of the QD. Lower Panel: Schematic representation of stadium BLG QD. **(c, d)** Numerical tight-binding simulations of electronic local density of states for **(c)** MLG and **(d)** BLG stadium QDs. $dI/dV_S$ diagonal streaks are visible in **(d)** (which has gapped barrier walls) but not in **(c)** (which has ungapped walls). $\gamma_3$ hopping and a spatially uniform 60 meV gap is included in the TB model of the BLG stadium. These parameters are motivated by our previous experimental characterization of circular BLG QDs [9], (also see SI section 6). The LDOS contribution from only sublattice $A_1$ is considered in the BLG stadium $dI/dV_s$ map simulation.

We are able to gain new insight into why scarred wavefunctions are absent in the QD stadia by examining the nature of the QD walls. For the MLG stadium, hole states inside of the QD are spatially adjacent to empty conduction-band electron states immediately outside of the stadium, thus leading to charge-carrier escape via Klein tunneling [26]. The resulting lack of strong confinement in MLG structures precludes the interference of unstable classical orbits that gives rise to scarring. This reasoning, however, suggests that scarred wavefunctions should be visible in a BLG stadium QD since the nature of the walls in a BLG stadium enable stronger confinement, thus increasing the interference between stable classical orbits. Our experimental findings for the BLG stadium QD, however, do not agree with this expectation.

To understand the absence of scarred wavefunctions in BLG stadium QDs we examine key differences between the QDs realized in our experiment and previously studied systems. In our experimental BLG QDs the nature of the confinement potential is smooth while in the initial theoretical work by Heller the confinement potential was sharp [13]. Smooth potential barriers, on the other hand, have been found to suppress quantum chaos in semiconductor-based QDs [27]. Such behavior has been attributed to dispersion in the energy of confined states at the boundaries which reduces the interference and mixing of states [27]. Stadiums with sharp potentials do not exhibit dispersion at the boundaries, thus allowing better interference and mixing of states that have similar energies and ultimately leading to wavefunction scarring [13].

To further support our hypothesis that the potential well sharpness precludes us from observing scarred wavefunctions for a BLG stadium in experiments, we simulated LDOS maps for a stadium-shaped BLG QD with a step potential well. Consequently, we found several possible scarred wavefunctions (data are shown in SI section 7). Video clips showing the evolution of simulated LDOS maps at different energies for a BLG stadium with a smooth potential well (in which scarred wavefunctions are absent) and a step potential well (in which candidate scarred wavefunctions are present) can also be found in the supporting information.

Additionally, we performed level statistical analysis for electrostatically defined MLG and BLG stadia with different potential well depth and sharpness (see SI section 9). Level statistical analysis has been widely used to investigate the chaotic behavior of either relativistic or non-relativistic quantum systems [22, 28]. We observed a transition from Poisson level-spacing distribution to Gaussian orthogonal ensemble distribution when the depth and sharpness of the potential well is increased for both MLG and BLG stadia. This finding indicates a deeper and sharper potential well enhances the chaotic behavior of both MLG and BLG stadia. For MLG stadia, we believe the chaotic behavior is enhanced with a deep and sharp potential well because of the onset of strong intervalley scattering caused by the sharp potential well. Such intervalley scattering will suppress Klein tunneling, which is a single-valley phenomenon.

Despite the absence of wavefunction scarring in our BLG stadium QDs, the appearance of diagonal streaks in both the experimental $dI/dV_S$ maps and the simulated BLG QD maps provides a connection to quantum chaotic phenomenon (additional simulated $dI/dV_S$ maps for a BLG stadium with different aspect ratios are provided in supporting information section 4). First, the absence of diagonal streaks in integrable systems regardless of confinement strength suggests that they are a special feature of nonintegrable systems [11, 13]. Second, we find that the streaks are absent in nonintegrable structures with poor confinement, such as MLG stadium QDs. This suggests that diagonal streaks require interference between states to form, similar to the phenomenon of wavefunction scarring.

In conclusion, we have fabricated and imaged electrostatically-defined stadium-shaped QDs in MLG and BLG sheets. Our wavefunction maps for these QDs reveal novel features (such as diagonal streaks) that originate from fundamental differences between the electronic structures of MLG and BLG. The absence of wavefunction scarring in MLG stadium QDs is attributed to charge carrier escape via Klein tunneling, while for BLG stadium QDs it is attributed to the smoothness of the confinement potential. These issues can potentially be addressed by placing BLG closer to the gating source through the use of thinner hBN. Such a change in the device architecture would also sharpen the electrostatic potential [29, 30], thus potentially enabling the interference between states that is necessary for wavefunction scarring.

**Supporting Information:**

(1) Numerical tight-binding calculation for electronic states within stadium-shaped MLG and BLG QDs, (2) $C_3$-symmetrical patterns at the end of stadium-shaped MLG and BLG QDs, (3) simulated $dI/dV_S$ patterns in gapped elliptical MLG and BLG QDs, (4) simulated $dI/dV_S$ maps for a stadium-shaped BLG QD with a different aspect ratio, (5) additional experimental $dI/dV_S$ maps for MLG and BLG stadia, (6) spatially resolved band gap for the experimental BLG stadium, (7) candidate scarred wavefunctions in a simulated BLG stadium with a step potential well, (8) level statistical analysis of an experimental BLG stadium, (9) level statistical analysis of simulated MLG and BLG stadia.

**Author contributions:** J.V.J., M.F.C., D.W., J.L., Z.G., and F.J., conceived the work and designed the research strategy. S.K., H.Z., Z.G., E.Q. fabricated the samples under A.Z. and J.V.J.'s supervision. K.W. and T.T. provided the hBN crystals. D.W., J.L., Z.G., F.J., and E.Q. carried out tunneling spectroscopy measurements under M.F.C. and J.V.J.'s supervision. Z.G. performed numerical tight-binding calculations and simulations with input from J.V.J. J.V.J., M.F.C., Z.G., and D.W. wrote the paper. All authors discussed the paper and commented on the manuscript.

**Acknowledgments:** This research was primarily supported by the sp2 program (KC2207) (STM imaging, device design) funded by the Director, Office of Science, Office of Basic Energy Sciences, Materials Sciences and Engineering Division, of the U.S. Department of Energy under Contract No. DE-AC02-05CH11231. Support was also provided by National Science Foundation grants no. DMR-1753367 (device fabrication, simulations, STM spectroscopy) and 1807233 (graphene growth), as well as the Army Research Office under contract W911NF-17-1-0473 (device characterization). Preparation and characterization of the BN crystals (K.W. and T.T.) was funded by the Elemental Strategy Initiative conducted by the MEXT, Japan, Grant Number JPMXP0112101001, JSPS KAKENHI Grant Number JP20H00354. We thank the Hummingbird Computational Cluster team at UC Santa Cruz for providing computational resources and support for the numerical tight-binding calculations performed in this work.

# Supporting Information

# Imaging Quantum Interference in Stadium-Shaped Monolayer and Bilayer Graphene Quantum Dots

Zhehao Ge[6,†], Dillon Wong[1,†], Juwon Lee[1], Frederic Joucken[6], Eberth A. Quezada-Lopez[6], Salman Kahn[1], Hsin-Zon Tsai[1], Takashi Taniguchi[4], Kenji Watanabe[4], Feng Wang[1,2,3], Alex Zettl[1,2,3], Michael F. Crommie[1,2,3,*], Jairo Velasco Jr.[6,*]

[1]*Department of Physics, University of California, Berkeley, California 94720, USA*
[2]*Materials Sciences Division, Lawrence Berkeley National Laboratory, Berkeley, California 94720, USA*
[3]*Kavli Energy NanoSciences Institute at the University of California, Berkeley and the Lawrence Berkeley National Laboratory, Berkeley, California 94720, USA*
[4]*International Center for Materials Nanoarchitectonics, National Institute for Materials Science, 1-1 Namiki, Tsukuba, 305-0044, Japan*
[5]*Research Center for Functional Materials, National Institute for Materials Science, 1-1 Namiki, Tsukuba, 305-0044, Japan*
[6] *Department of Physics, University of California, Santa Cruz, California 95064, USA*
[†] *These authors contribute equally to this manuscript.*
*\* Email: jvelasc5@ucsc.edu & crommie@berkeley.edu*

**Table of Contents**

**S1. Numerical tight-binding calculation for electronic states within stadium-shaped MLG and BLG QDs**

**Stadium-shaped MLG QD LDOS calculation**

The stadium-shaped MLG QD is modeled by the following tight-binding (TB) Hamiltonian within the first-nearest-neighbor-hopping approximation on a circular MLG sheet with a 450 nm radius:

$$H = \sum_i [V(\vec{R}_i^A) + \frac{\Delta}{2}]a_i^\dagger a_i + \sum_i [V(\vec{R}_i^B) - \frac{\Delta}{2}]b_i^\dagger b_i - \sum_{<i,j>} \gamma_0(a_i^\dagger b_j + b_j^\dagger a_i)$$

where the operators $a_i^\dagger (a_i)$ and $b_i^\dagger (b_i)$ create (annihilate) an electron on site $\vec{R}_i$ of sublattice $A$ and $B$, respectively. The stadium-shaped MLG QD is defined by varying the onsite energy $V$ of carbon atoms at the position $\vec{r} = \vec{R}_i$. The $V(\vec{r})$ used for simulating stadium-shaped MLG p-n junctions are shown in Figs. S1a and S1b. The hopping parameter between the nearest carbon atoms is $\gamma_0$, where we used $\gamma_0 = 3.3\ eV$. This corresponds to a Fermi velocity $v_F \approx 1.07 \times 10^6\ m/s$ for graphene. The local density of states $LDOS(E,r)$ was computed numerically from the above Hamiltonian using the Pybinding package [1], which uses the kernel polynomial method [2] to solve the Hamiltonian.

**Stadium-shaped BLG QD LDOS calculation**

The stadium-shaped BLG QD is modeled by the following tight-binding (TB) Hamiltonian within the first-nearest-neighbor-hopping approximation on a circular BLG sheet with a 250 nm radius:

$$\begin{aligned}H &= \sum_i [V(R_i^{A1}) + \frac{U}{2}]a_{1i}^\dagger a_{1i} + \sum_i [V(R_i^{B1}) + \frac{U}{2}]b_{1i}^\dagger b_{1i} \\ &+ \sum_i [V(R_i^{A2}) - \frac{U}{2}]a_{2i}^\dagger a_{2i} + \sum_i [V(R_i^{B2}) - \frac{U}{2}]b_{2i}^\dagger b_{2i} - \sum_{<i,j>} \gamma_0(a_{1i}^\dagger b_{1j} \\ &+ H.c.) - \sum_{<i,j>} \gamma_0(a_{2i}^\dagger b_{2j} + H.c.) + \sum_{<i,j>} \gamma_1(b_{1i}^\dagger a_{2j} + H.c.) \\ &- \sum_{<i,j>} \gamma_3(a_{1i}^\dagger b_{2j} + H.c.)\end{aligned}$$

where the operators $a_{1i}^\dagger(a_{1i})$, $b_{1i}^\dagger(b_{1i})$, $a_{2i}^\dagger(a_{2i})$ and $b_{2i}^\dagger(b_{2i})$ create (annihilate) an electron on site $\vec{R}_i$ of sublattice $A_1, B_1, A_2$ and $B_2$, respectively. The stadium-shaped BLG QD is defined by varying the onsite energy $V$ of the carbon atoms as a function of their position, which is denoted by $\vec{r} = \vec{R}_i$. The $V(\vec{r})$ used for simulating the stadium-shaped BLG p-n junctions are shown in Figs. S1c and S1d. The hopping parameters [3] used in the model are $\gamma_0 = 3.16$ eV, $\gamma_1 = 0.381$ eV and $\gamma_3 = -0.38$ eV. For gapped BLG QDs, the interlayer potential difference is chosen as $U = 60\ meV$. The local density of states $LDOS(E, r)$ was computed numerically from the above Hamiltonian using the Pybinding package [1], which uses the kernel polynomial method [2] to solve the Hamiltonian. In all our BLG stadia LDOS map simulations, only the LDOS contribution from sublattice $A_1$ is considered because the QD states are polarized on this top layer sublattice. In addition, this top layer sublattice contributes most of the $dI/dV_S$ signal in our STM $dI/dV_S$ map measurements.

**$dI/dV_S$ simulation**

In our experiments, we use a constant current $I$ at a certain sample bias voltage $V_S$ to control the distance between the STM tip and sample. But since a QD's LDOS is position dependent, the distance between the STM tip and sample can vary at different locations of the QD. This is because the tunneling current $I \propto |e^{-\alpha z} \int_0^{eV_S} LDOS(E, r) dE|$ is a constant [4], which causes $e^{-\alpha z} \propto \frac{I}{|\int_0^{eV_S} LDOS(E,r)dE|}$ to vary, hence the tip-sample distance changes. Since $\frac{dI}{dV_S} \propto e^{-\alpha z} LDOS(E, r)$, we can simulate the tip-sample distance variation effect by using the following formula to calculate $dI/dV_S$ based on the calculated $LDOS(E, r)$ for graphene p-n junctions:

$$\frac{dI}{dV_S}(E, r) \propto \frac{LDOS(E, r)}{|\int_0^{eV_S} LDOS(E, r) dE|}$$

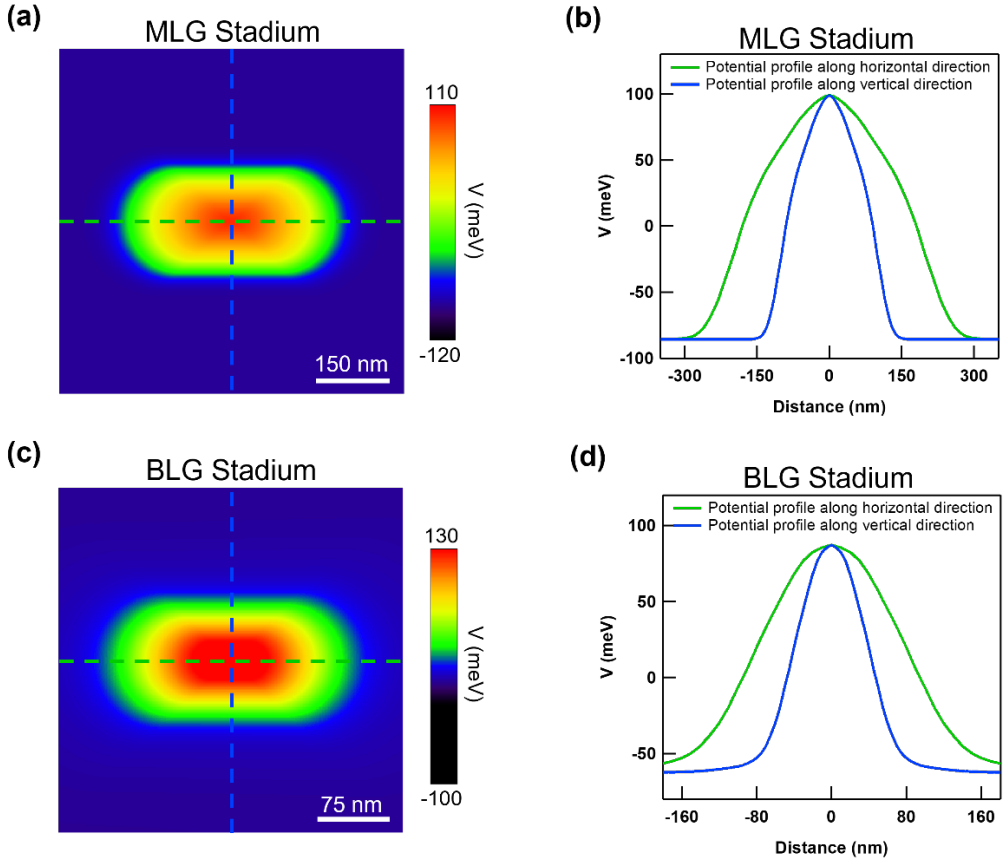

**Figure S1: Potential wells $V(\vec{r})$ used for simulated stadium-shaped MLG and BLG QDs**. (**a**) Spatial color map depicting the onsite energy of graphene carbon atoms in the TB model. This defines $V(\vec{r})$ for the stadium-shaped MLG QD. (**b**) Profiles of the stadium-shaped MLG QD potential $V(\vec{r})$ from **a**. The profiles are along the horizontal and vertical directions and correspond to the green and blue dashed lines in **a**, respectively. (**c**) Spatial color map depicting the onsite energy of BLG carbon atoms in the TB model. This defines $V(\vec{r})$ for the stadium-shaped BLG QD. (**d**) Profiles of the stadium-shaped BLG QD potential $V(\vec{r})$ from **c**. The profiles are along the horizontal and vertical directions which are depicted by the green and blue dashed lines in **c**, respectively.

**S2. $C_3$-symmetric patterns at the end of stadium-shaped MLG and BLG QDs**

Additional $dI/dV_S$ map simulations from the same stadium-shaped MLG and BLG QDs as presented in Fig. 4 are shown in Fig. S2. These additional results correspond to QD states with lower energies compared to the QD states shown in Fig. 4 and more clearly reveal the C3-symmetric patterns near the end of stadium-shaped MLG and BLG QDs. As can be seen in Fig. S2c the orientation of the triangular $dI/dV_S$ patterns in MLG stadiums are reversed at the two ends and point outward. In contrast, as shown in Fig. S2d, the orientation of the triangular $dI/dV_S$ patterns in BLG stadiums are aligned at the two ends, with one pointing outward and the other pointing inward.

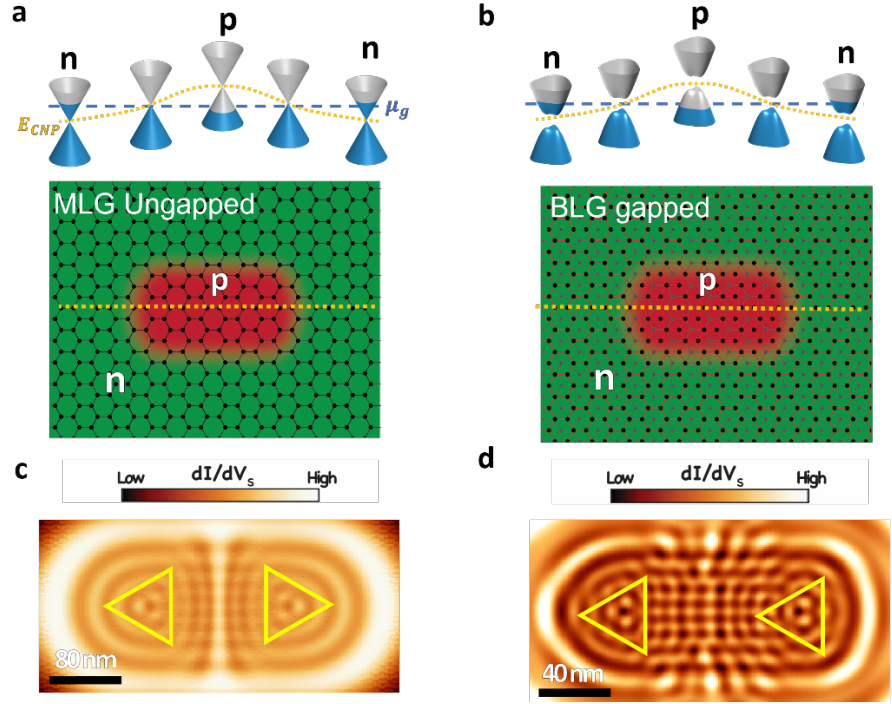

**Figure S2: Triangular $dI/dV_S$ patterns at the end of stadium-shaped MLG and BLG QDs.** **(a)** Upper panel: Stadium MLG QD potential profile schematic depicting the MLG bands and charge neutrality ($E_{CNP}$) within and outside of the QD. Lower Panel: Schematic representation of stadium-shaped MLG QD. **(b)** Upper panel: Stadium BLG QD potential profile schematic depicting the gapped and trigonally warped BLG bands and $E_{CNP}$ within and outside of the QD. Lower Panel: Schematic representation of stadium-shaped BLG QD. **(c-d)** Numerical tight-binding (TB) simulations of $dI/dV_S$ maps for ungapped MLG and gapped BLG stadium-shaped QDs, respectively. The energies of the QD states in **c-d** are lower than that of the QD states shown in Fig. 4. The yellow triangles in **c-d** depict the orientations of the triangular patterns at the end of the stadium-shaped QDs. The LDOS contribution from only sublattice $A_1$ is considered in the BLG stadium $dI/dV_S$ map simulation.

## S3. Simulated $dI/dV_S$ patterns in gapped elliptical BLG QD

Simulated $dI/dV_S$ maps for elliptical gapped BLG QDs are shown in Fig. S3b. The faint diagonal d$I/dV_S$ streaks that appeared in the stadium-shaped BLG QDs are absent in Figs. S3b.

This result indicates that the faint diagonal $dI/dV_S$ streaks in stadium-shaped BLG QDs are related to the non-integrability of the stadium-shaped QD.

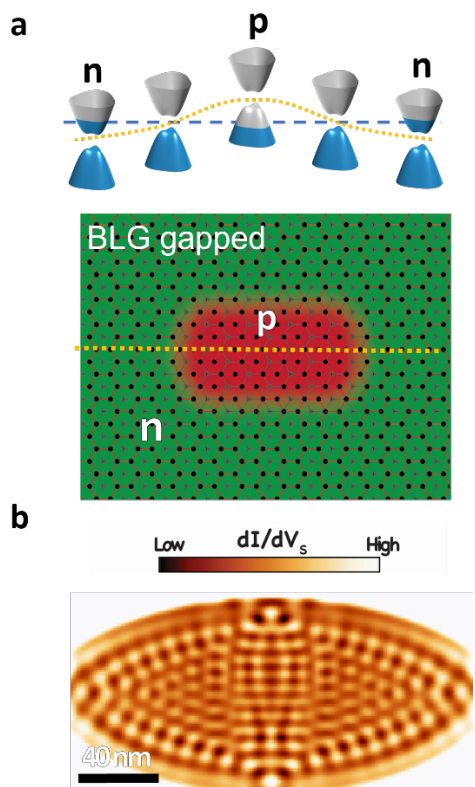

**Figure S3: Simulated $dI/dV_S$ patterns for elliptical BLG QDs. (a)** Upper panel: Elliptical BLG QD potential profile schematic depicting the gapped and trigonally warped BLG bands and $E_{CNP}$ within and outside of the QD. Lower Panel: Schematic representation of elliptical BLG QD. **(b)** Numerical tight-binding (TB) simulation of $dI/dV_S$ map for gapped elliptical BLG QD. The diagonal $dI/dV_S$ streaks present in stadium-shaped BLG QD in Figs. 4d is absent in the elliptical BLG QD. The LDOS contribution from only sublattice $A_1$ is considered in the BLG stadium $dI/dV_S$ map simulation.

**S4. Simulated $dI/dV_S$ maps for a stadium-shaped BLG QD with a different aspect ratio**

Simulated $dI/dV_S$ maps for a stadium-shaped BLG QD with a different aspect ratio are shown in Fig. S4. The diagonal streaks reported in the main text of the manuscript also appear here. This indicates that the observed diagonal streaks are not due an elongated confinement structure.

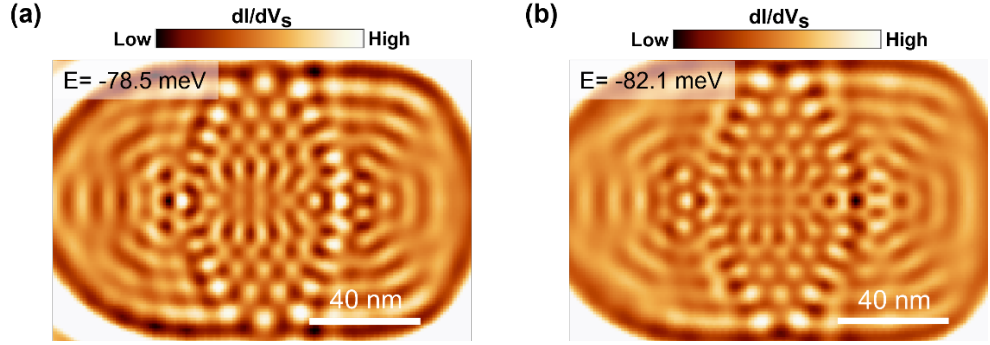

**Figure S4: Simulated $dI/dV_S$ maps for a less elongated BLG stadium. a-b,** Simulated $dI/dV_S$ maps for an electrostatically defined BLG QD at $E = -78.5$ meV and $E = -82.1$ meV, respectively. $\gamma_3$ hopping and a spatially uniform 60 meV gap is included in the TB model, and $V_S = -15$ meV is used for $dI/dV_S$ simulation. The LDOS contribution from only sublattice $A_1$ is considered in the BLG stadium $dI/dV_S$ map simulation.

## S5. Additional experimental $dI/dV_S$ maps for MLG and BLG stadia

Figure S5 shows additional experimental $dI/dV_S$ maps taken at various $V_G$ but with the same $V_S$ for the same stadium-shaped BLG QD shown in the main text. Modulating $V_G$ while $V_S$ is maintained constant is nearly equivalent to changing $V_S$ at a constant $V_G$ to probe QD states with different energies. There is no clear scarred wavefunctions observed in these data. Figure S6 shows additional experimental $dI/dV_S$ maps taken at various $V_G$ but with the same $V_S$ for the same stadium-shaped MLG QD shown in the main text. Here there is also no clear evidence for scarred wavefunctions.

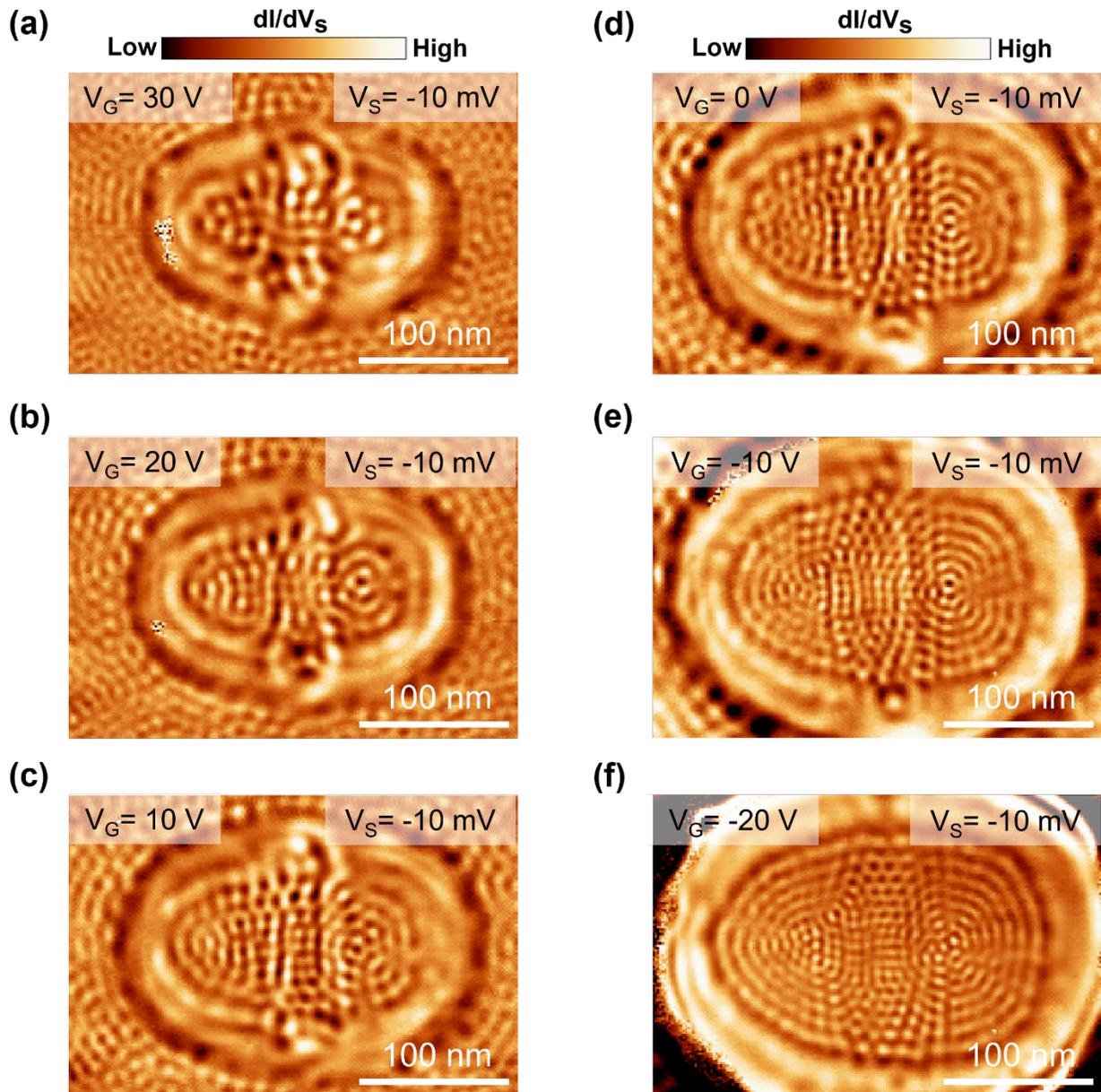

**Figure S5: The absence of scarred wavefunctions in an experimental BLG stadium. a-f,** Experimental $dI/dV_S$ maps measured at various $V_G$ for a stadium-shaped BLG QD. No scarred wavefunctions are observed for the BLG stadium with the gate and bias values shown here. The scanning parameters used to acquire these images were $I = 0.1$ nA, $V_S = -10$ mV and with a 2 mV ac excitation.

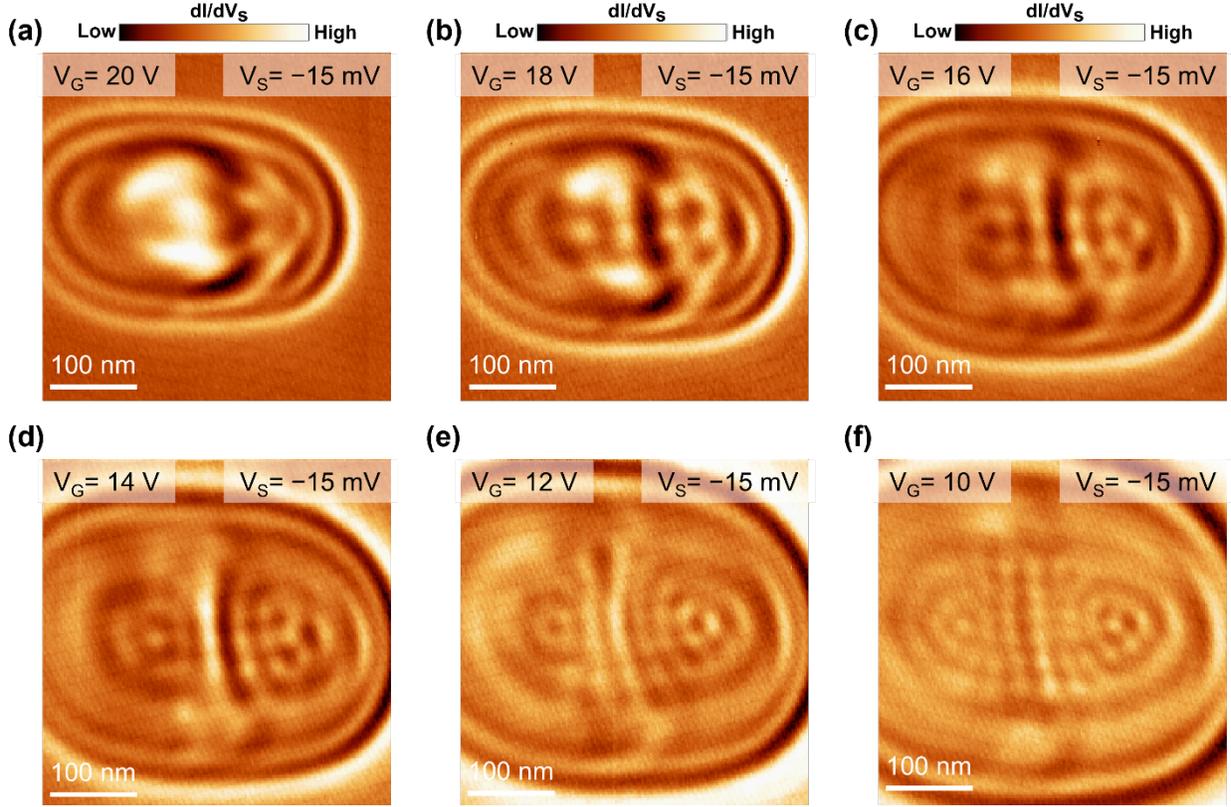

**Figure S6: The absence of scarred wavefunctions in an experimental MLG stadium. a-f,** Experimental $dI/dV_S$ maps measured at various $V_G$ for a stadium-shaped MLG QD. No scarred wavefunctions are observed for the MLG stadium with the gate and bias values shown here. The scanning parameters used to acquire these images were $I = 0.5$ nA, $V_S = -15$ mV and with a 2 mV ac excitation.

## S6. Spatially resolved band gap for the experimental BLG stadium

Figure S7a shows one experimentally measured $dI/dV_S(V_S, d)$ across the center of a stadium-shaped BLG QD, Fig. S7b shows the $dI/dV_S$ spectra at $d = -100$ nm. A gap value around 60 meV that has small spatial variation is observed. Additionally, this value is close to the gap value we observe for circular BLG QDs that are created by the same hBN defect ionization method[5] (related circular BLG QD data can be found in the supplementary section 1 of reference [5]).

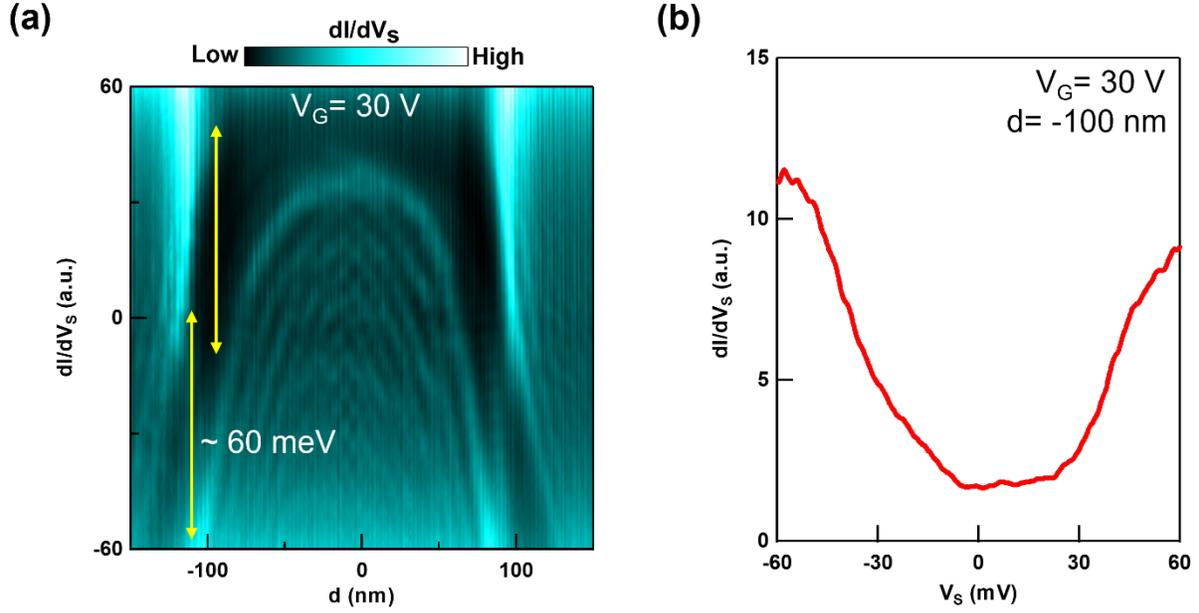

**Figure S7: Spatially resolved $dI/dV_S$ spectra for a BLG stadium. a,** Experimentally measured $dI/dV_S(V_S, d)$ at $V_G = 30$ V for a stadium-shaped BLG QD along a line that crosses the stadium center and along the long direction of the stadium. This BLG stadium is not the same one as shown in the main text, though a similar fabrication procedure was used for both stadia. The yellow arrow indicates the gap size of the BLG stadium. The set point used to acquire the tunneling spectra was $I = 1$ nA, $V_S = -60$ mV, with a 2 mV ac modulation. **b,** $dI/dV_S$ spectrum at $d = -100$ nm from (a).

## S7. Candidate scarred wavefunctions in a simulated BLG stadium with a step potential well

Figure S8 shows the simulated LDOS maps of some candidate scarred wavefunctions for a BLG stadium with a step potential well. Video clips showing the evolution of simulated LDOS maps at different energies can also be found in the supporting information.

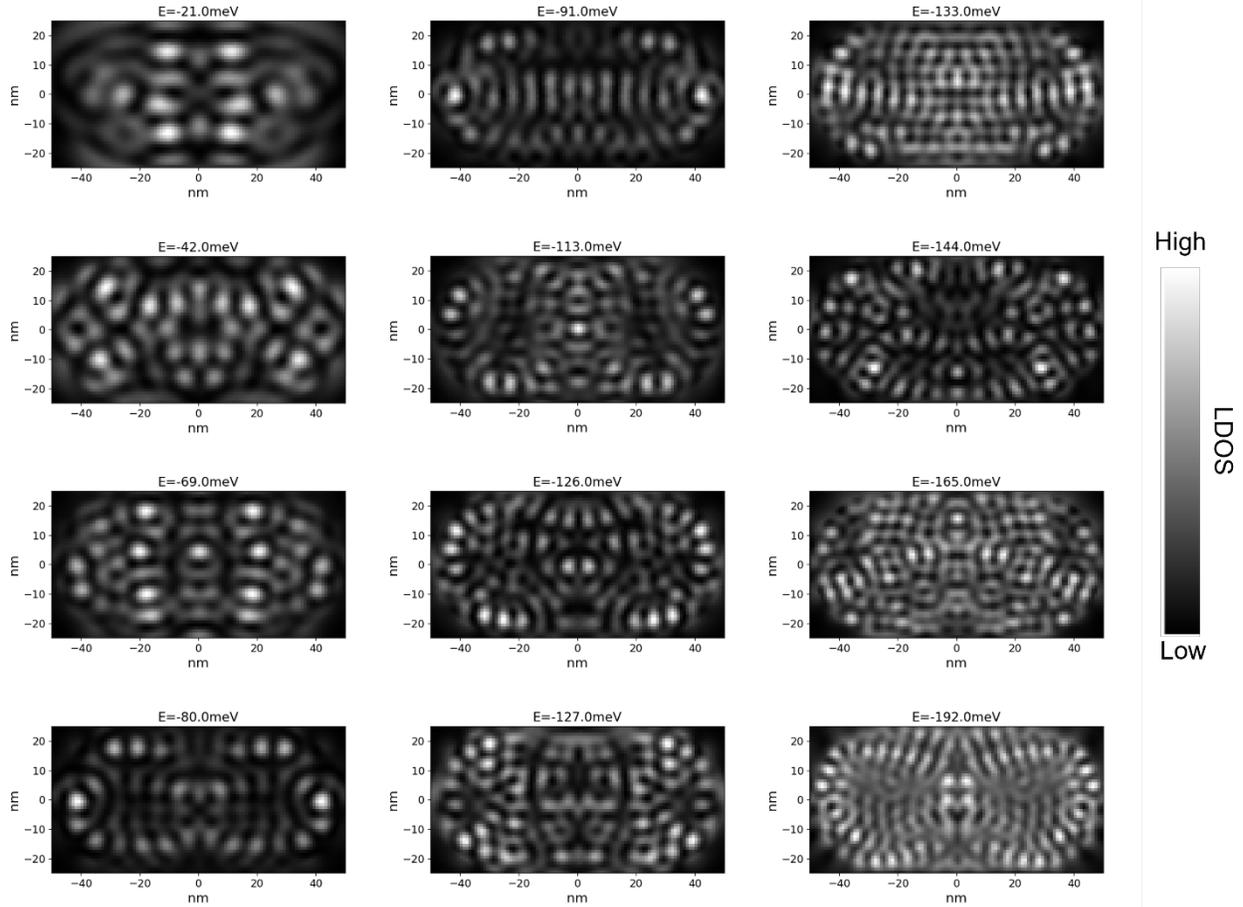

**Figure S8: Candidate scarred wavefunctions in a BLG stadium with a step potential well.** Simulated LDOS maps at some selected energies that resemble scarred wavefunctions for an electrostatically defined BLG stadium with a step potential well. The depth of the step potential well is 200 meV. The BLG stadium is composed of a $50 \times 50$ nm$^2$ square at the center and two semicircles that are connected at the ends. The $\gamma_3$ hopping is ignored and a spatially uniform 200 meV interlayer energy difference is included in the TB model. The LDOS contribution from only sublattice $A_1$ is considered in the BLG stadium LDOS map simulation.

**S8. Level statistical analysis of an experimental BLG stadium**

In our experiments the number of observed LDOS peaks are too small, thus preventing a reliable energy-level statistical analysis. Figure S9a shows an example $dI/dV_S$ spectra measured at the center of a stadium-shaped BLG QD. By tracking the local maximum of the $dI/dV_S$ spectrum, we extract the energies of different QD states. The corresponding level-spacing distribution of the extracted states is plotted in Fig. S9b. This result does not agree with a Poisson or a Gaussian orthogonal ensemble (GOE) distribution.

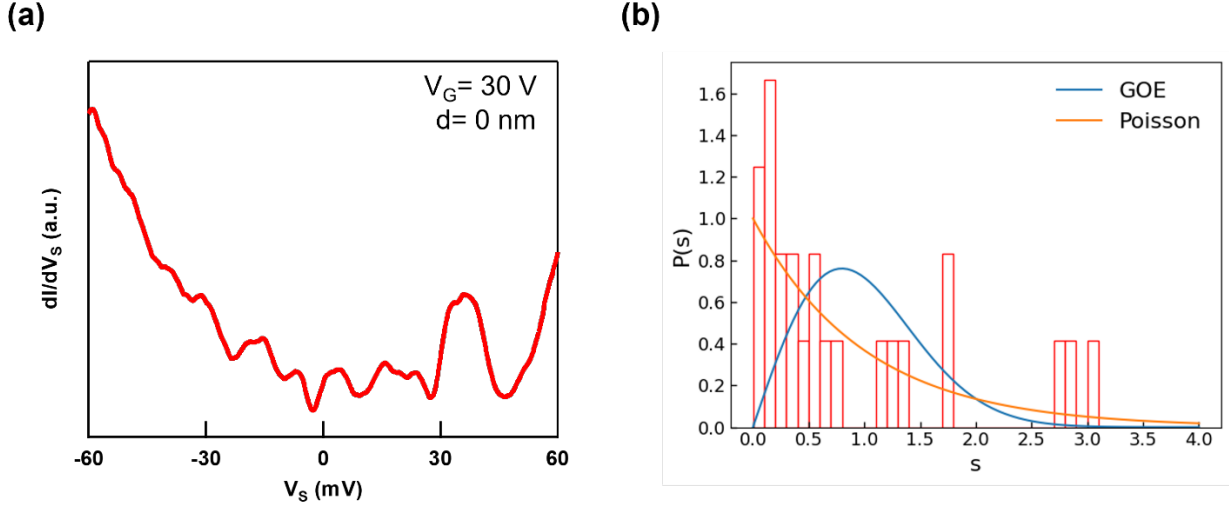

**Figure S9:** $dI/dV_S$ **spectrum and level statistics of a BLG stadium a,** $dI/dV_S$ spectrum taken at the center of a stadium-shaped BLG QD at $V_G = 30$ V. The set point used to acquire the tunneling spectra was $I = 1$ nA, $V_S = -60$ mV, with a 2 mV ac modulation. **b,** $dI/dV_S$ peak spacing distribution for the peaks extracted from (a), the peak spacing are normalized as $s = \frac{\Delta V_S}{<\Delta V_S>}$.

## S9. Level statistical analysis of simulated MLG and BLG stadia

**Method**

The TB Hamiltonians of MLG and BLG nano sheets with different physical shapes and electrostatic potential wells are first defined similar to as described in supporting information section1. Next, the eigenvalues of the Hamiltonian are solved by the LAPACK eigenvalue solver implemented in the Pybinding package [1]. Then, the eigenvalues are ordered and indexed in sequence from low energy to high energy. Finally, we select a portion of the sequenced eigenvalues to perform level statistical analysis with a standard procedure as described in ref [6]. We noticed the level statistical results are weakly dependent on the index range used to perform the level statistical analysis. We chose an index range of 50%-70% for the level statistics analysis because it clearly yields results that are consistent with prior works [6, 7].

**Level statistical results for physically defined MLG and BLG stadia**

Figures S10a-c show the structure and size of MLG QDs used in the TB calculation with stadium, rectangular, and circular shapes, respectively. Figures S10d-f show the energy of the calculated states that fall in the 50% to 70% index range of the full sequence of ordered eigenvalues calculated from the corresponding structures. Figures S10g-i shows the level-spacing distribution of the states shown in Figures S10d-f, the level spacings are normalized as $\frac{\Delta E}{<\Delta E>}$. For stadium-shaped MLG QDs the level spacing agrees very well with the GOE distribution (Fig. S10g), which is expected for non-integrable systems [6, 7]. For rectangular MLG QDs, the level-spacing distribution displayed good agreement with the Poisson distribution (Fig. S10h), which is expected for integrable systems [6, 7]. Similar results are observed for gapped BLG QDs, the level-spacing distribution for rectangular- and stadium-shaped BLG QD shows a Poisson distribution and a GOE distribution, respectively. These results are plotted in Fig. S12.

For the circular MLG QD, which is considered an integrable system, the expectation is that its level-spacing distribution should agree with the Poisson distribution. However, we found that the level-spacing distribution for the MLG QD does not agree with either the Poisson or the GOE distribution (Fig. S10i). We attribute this behavior to the deviations from a circular confinement structure for the circular MLG nano sheet defined in the TB model. As can be seen more clearly in Fig. S11, the MLG QD boundary is not exactly circular, hence this could possibly lead to a mixture of Poisson and GOE distributions.

**Level statistical results for electrostatically defined MLG and BLG stadia**

Besides physically defined stadium-shaped QDs, we also investigated electrostatically defined stadium-shaped QDs in rectangular MLG and BLG sheets. These simulations are more relevant to our experiment. As shown in Fig. S13, we observed the level-spacing distribution for a stadium-shaped MLG QD transitioned from a Poisson distribution to a GOE distribution as the depth and sharpness of the QD potential well increased. A similar result is achieved for the stadium-shaped BLG QD (Fig. S14). These results suggest a deeper and shaper potential well enhances the chaotic behavior of both stadium-shaped MLG and BLG QDs that are electrostatically defined.

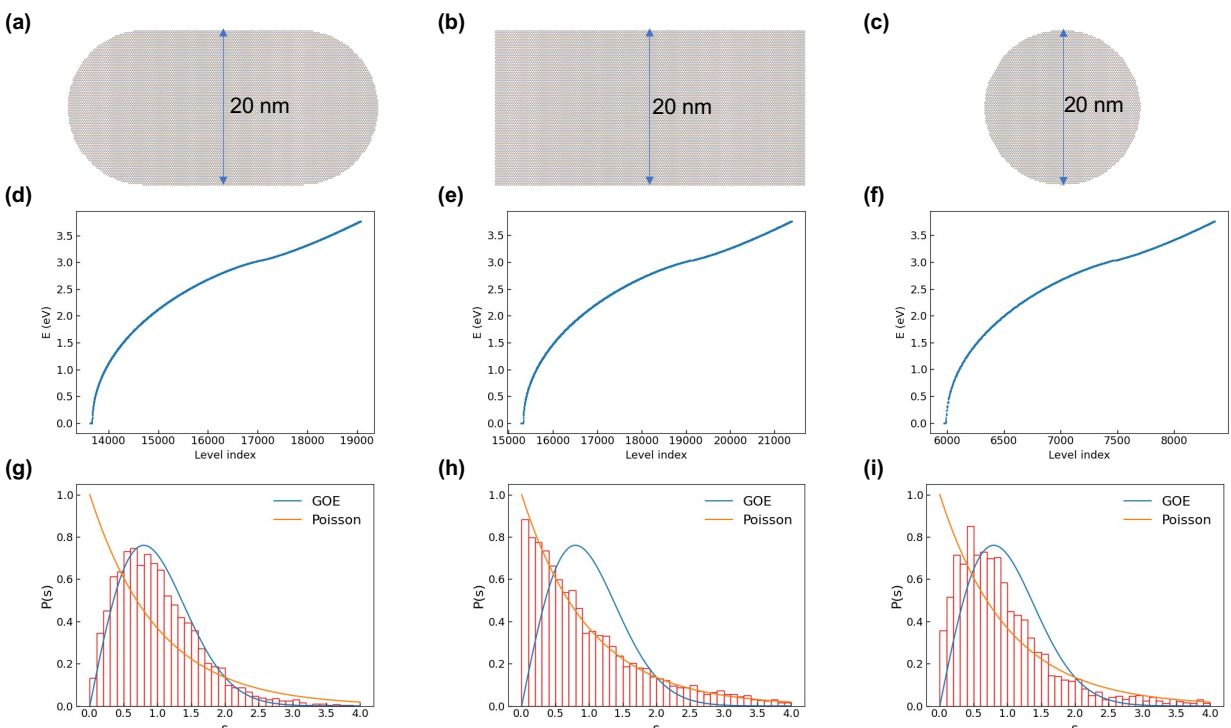

**Figure S10: Physically defined MLG QDs with different shapes and their level statistics a-c,** The shape, and size of the MLG QD defined in our TB model. For (a) a stadium consisting of a square at the center connected by two semicircles at the two ends, for (b) a rectangle with a 1:2 ratio between the width and length, and for (c) a circle. **d-f,** Some low-energy eigenvalues calculated from the MLG QDs shown in (a)-(c), respectively. The x axis represents the sequence of the eigenvalues that are ordered from low energy to high energy. **g-i,** The level-spacing distribution of the eigenvalues shown in (d)-(f), respectively. The level spacings are normalized as $s = \frac{\Delta E}{<\Delta E>}$.

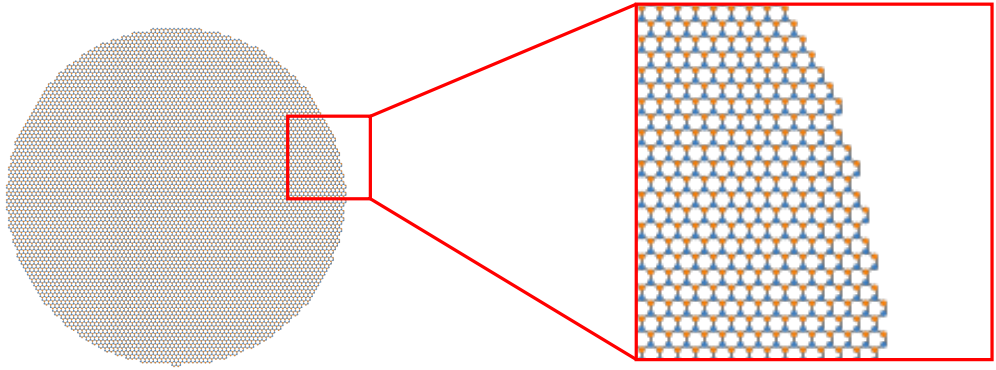

**Figure S11: Atomic-scale irregularities at the boundary of a circular MLG QDs.** The left panel shows the shape of the same MLG QD as shown in Fig. R2c. The right panel is the zoom in of a boundary for this circular MLG QD, irregular edges can be seen.

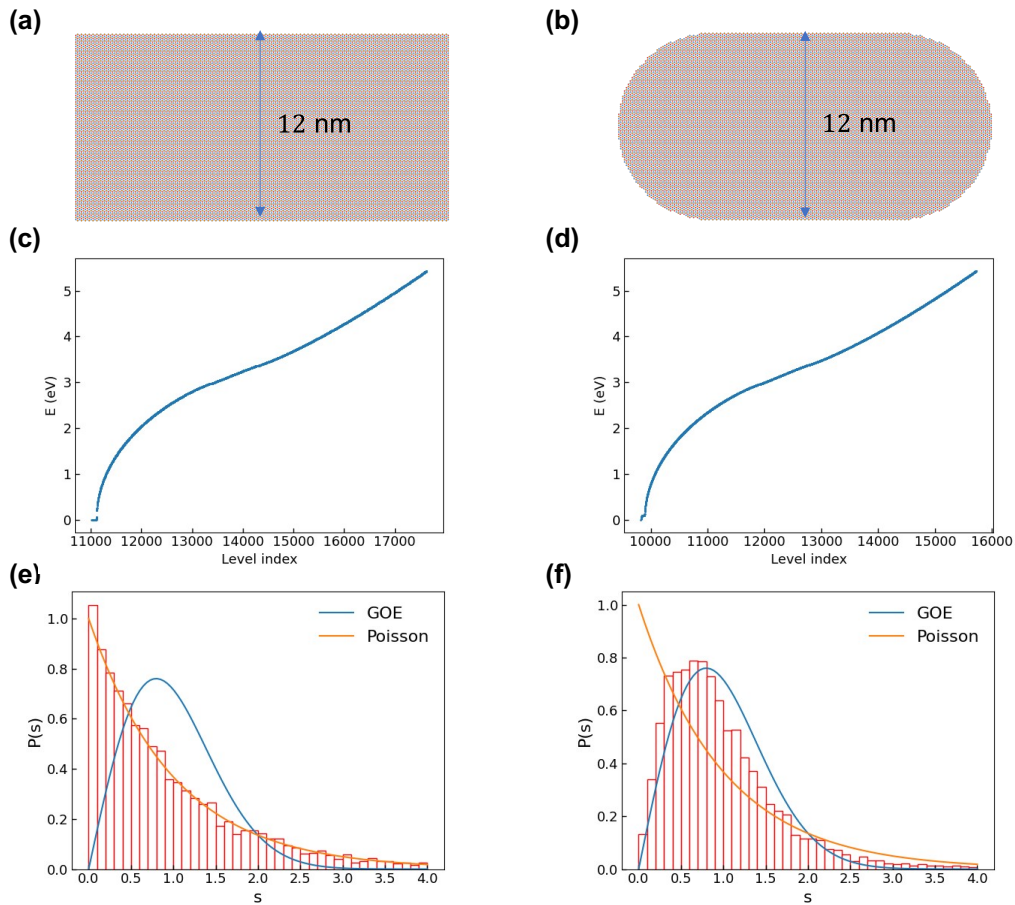

**Figure S12: Physically defined BLG QD with different shapes and their level statistics. a-b,** The shape, and size of the BLG QD defined in the TB model. For (a) a rectangle with a 1:2 ratio between the width and length, for (b) a stadium consisting of a square at the center connected by two semicircles at the two ends. $\gamma_3$ hopping is ignored and a 200 meV interlayer potential difference is used in the TB model for both (a) and (b). **c-d,** Some low-energy eigenvalues

calculated from the BLG QDs shown in (a)-(b), respectively. The x axis represents the sequence of the eigenvalues that are ordered from low energy to high energy. **e-f,** The level-spacing distribution of the eigenvalues shown in (c)-(d), respectively. The level spacings are normalized as $s = \frac{\Delta E}{<\Delta E>}$.

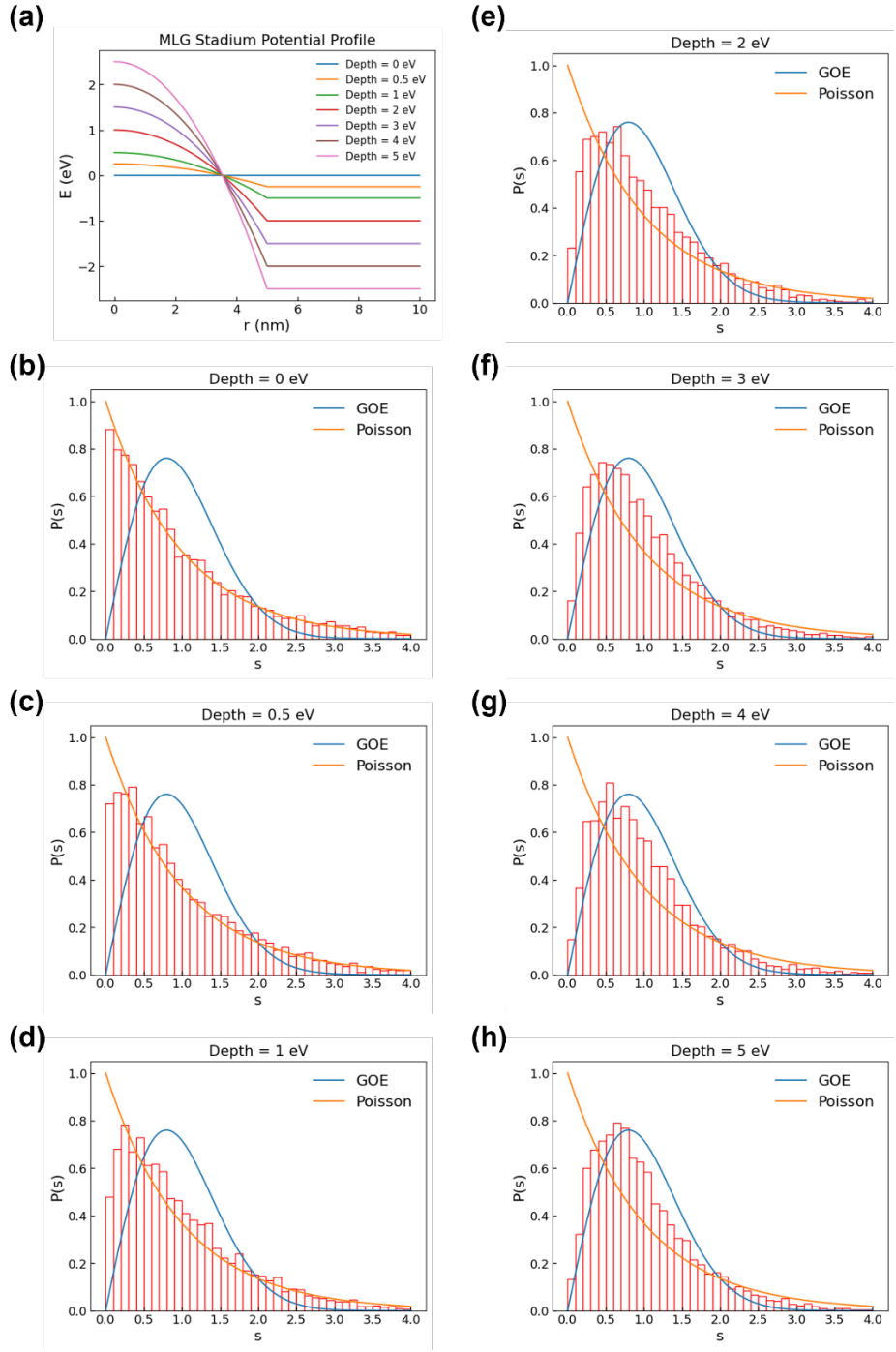

**Figure S13: Level statistics of electrostatically defined MLG stadia with different potential well depth and sharpness. a,** Potential profiles of electrostatically defined MLG stadium in a rectangular $20 \times 40$ nm² MLG sheet along the shorter direction of the rectangle. Between $r = 0$ nm and $r = 5$ nm, the potential profiles are defined by quadratic functions with different depth and sharpness. Between $r = 5$ nm and $r = 10$ nm, the potential profiles are flat. **b-h,** The level-spacing distribution of the eigenvalues for MLG stadia with different potential depth and sharpness as defined in (a). The eigenvalues in the 50% to 70% index range of the full sequence of ordered eigenvalues are used for statistical analysis. The level spacings are normalized as $s = \frac{\Delta E}{<\Delta E>}$.

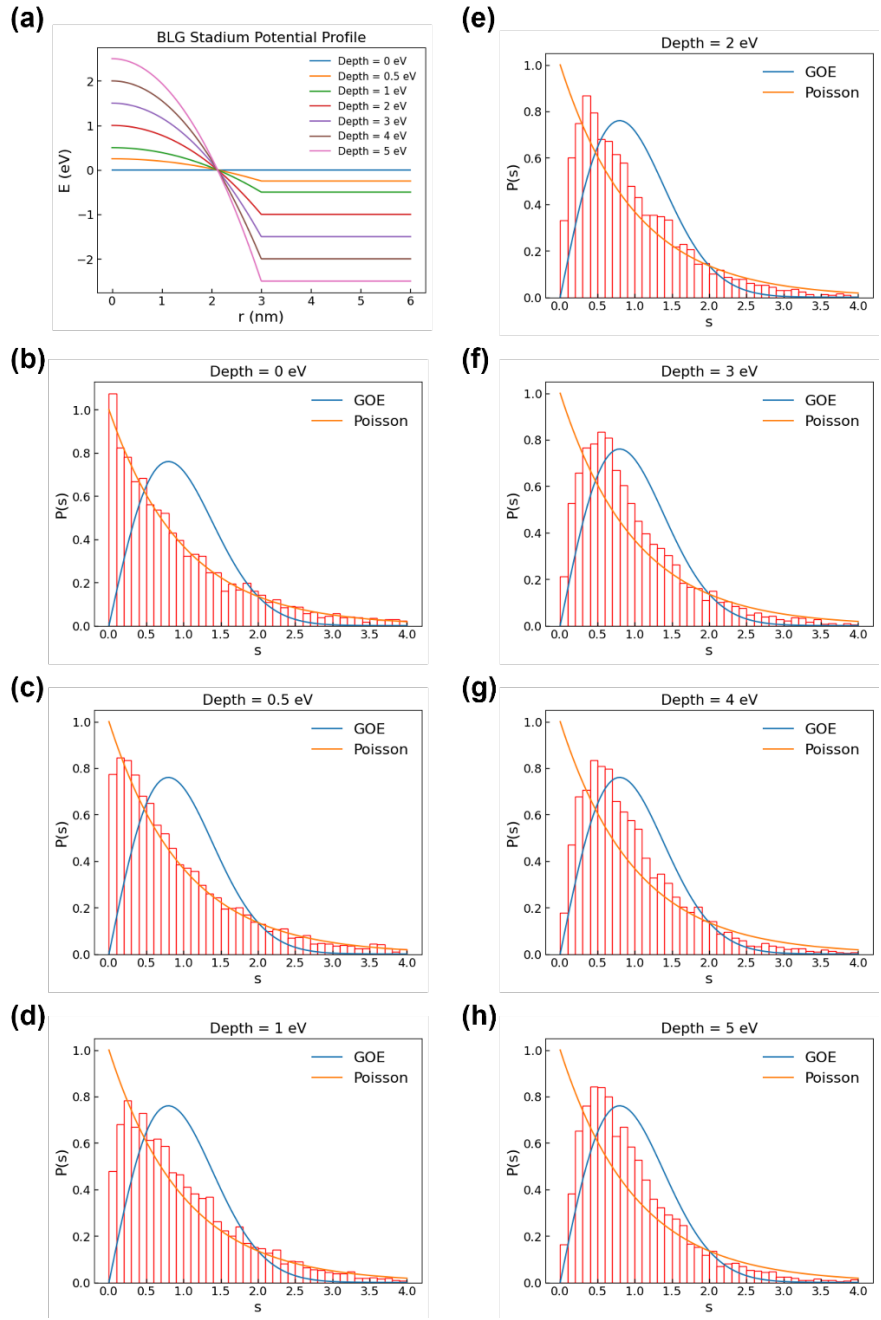

**Figure S14: Level statistics of electrostatically defined BLG stadia with different potential well depth and sharpness. a,** Potential profiles of electrostatically defined BLG stadium in a rectangular $12 \times 24$ nm$^2$ BLG sheet along the shorter direction of the rectangle. Between $r = 0$nm and $r = 3$nm, the potential profiles are defined by quadratic functions with different depth and sharpness. Between $r = 3$nm and $r = 6$nm, the potential profiles are flat. $\gamma_3$ hopping is ignored and a 200 meV interlayer potential difference is used in the model. **b-h,** The level-spacing distribution of the eigenvalues for MLG stadia with different potential depth and sharpness as defined in (a). The eigenvalues in the 50% to 70% index range of the full sequence of ordered eigenvalues are used for statistical analysis. The level spacings are normalized as $s = \frac{\Delta E}{<\Delta E>}$.